\documentclass[aps,prd,onecolumn,amssymb]{revtex4}
\usepackage{graphicx,bm,color}
\usepackage[english]{babel}
\usepackage{amsmath}
\usepackage{amsfonts}

\allowdisplaybreaks[4]

\newcommand{\be}{\begin{equation}}
\newcommand{\ee}{\end{equation}}
\newcommand{\bea}{\begin{eqnarray}}
\newcommand{\eea}{\end{eqnarray}}
\newcommand{\beaa}{\begin{eqnarray*}}
\newcommand{\eeaa}{\end{eqnarray*}}

\begin{document}

\title{Turbulence and Little Rip Cosmology}


\author{
I.~Brevik$^{a,}$\footnote{E-mail: iver.h.brevik@ntnu.no},
R.~Myrzakulov$^{b,}$\footnote{Email: rmyrzakulov@gmail.com; rmyrzakulov@csufresno.edu}
S.~Nojiri$^{c,d,}$\footnote{E-mail: nojiri@phys.nagoya-u.ac.jp}
and S.~D. Odintsov$^{e,f,b,g,}$\footnote{E-mail: odintsov@ieec.uab.es}
}

\affiliation{
$^a$ Department of Energy and Process Engineering,
Norwegian University of Science and Technology, N-7491 Trondheim, Norway \\
$^b$ Eurasian International Center for Theoretical Physics and Department of General
\& Theoretical Physics, Eurasian National University, Astana 010008, Kazakhstan \\
$^c$ Department of Physics, Nagoya University, Nagoya 464-8602, Japan \\
$^d$ Kobayashi-Maskawa Institute for the Origin of Particles and the Universe,
Nagoya University, Nagoya 464-8602, Japan, \\
$^e$Consejo Superior de Investigaciones Cient\'{\i}ficas, ICE/CSIC-IEEC,
Campus UAB, Facultat de Ci\`{e}ncies, Torre C5-Parell-2a pl, E-08193
Bellaterra (Barcelona) Spain \\
$^f$ Instituci\'{o} Catalana de Recerca i Estudis Avan\c{c}ats (ICREA),
Barcelona \\
$^g$ Tomsk State
Pedagogical Univ., Tomsk, Russia
}

\begin{abstract}

A variety of conditions is considered under which the cosmic dark fluid may be
able
to develop a future Big Rip or Little Rip singularity. Both
one-component and two-component models are considered. In the last-mentioned
case we present
a way in which the fluid can be decomposed into two components,
one non-turbulent (ideal) and one turbulent part, obeying two different
equations of state.
For the non-turbulent part, the thermodynamical parameter, commonly
called $w$, is assumed to be less than $-1$ throughout.
For the turbulent part, it turns out that it is sufficient that
$w_\mathrm{turb}$ lies in the
quintessence region in order to lead to a singularity. Both Big Rip and Little
Rip behaviour for dark energy are found.
In the one-component case, we examine how the universe
may develop from a viscous era with constant bulk viscosity into a turbulent
era, the turbulence in effect protecting the universe from encountering the
singularity at all.
The equivalent description of the same cosmology in terms of inhomogeneous
(imperfect) fluid is also presented.

\end{abstract}

\maketitle

\section{Introduction}

It has become customary to explain the observed acceleration of the universe
\cite{Riess,Perlmutter}
in terms of  dark energy fluid (for recent reviews, see
\cite{Dark-6,Cai:2009zp}), which is expected to have the
strange properties like negative pressure and/or negative entropy.
According to the latest supernovae observations the dark energy amounts to
about 73\% of the total mass energy of the universe \cite{Kowalski}.
Although astrophysical observations favor the standard $\Lambda$CDM cosmology,
the equation-of-state (EoS) parameter $w$ is still determined with
uncertainty: it is not clear if $w$ is less than $-1$, equal to $-1$, or
larger than $-1$.
Current observations suggest that  $w=-1.04^{+0.09}_{-0.10}$ \cite{PDP,Amman}.

The very interesting but least theoretically understood case corresponds to
$w<-1$ (phantom dark energy)  where all four energy conditions are violated.
Although the theory is unstable from a quantum field theoretical viewpoint,
it could be stable in classical cosmology.
There are observations \cite{P} indicating that the crossing of the
cosmological constant/phantom divide took  place
in the near past (or will occur in the near future). An essential property of
(most) phantom dark energy models
is the Big Rip future singularity \cite{caldwell02} (see also
\cite{BR,Nojiri}),
where the scale factor
becomes infinite at a finite time in the future. A softer future singularity
caused by phantom or quintessence dark energy is the sudden (Type II)
singularity \cite{Barrow}
where the scale factor is finite at Rip time (for classification of finite-time
singularities, see \cite{Nojiri-2}). Recently, an attempt to resolve the
finite-time future singularities has been proposed in the face of mild phantom
models where $w$ asymptotically tends to $-1$ and where the energy-density
increases with the time or remains constant, but where singularity occurs in
infinite future\cite{Frampton-2,Frampton-3,Astashenok,others}. The key point
here
is that if $w$ approaches $-1$ sufficiently fast, then it is possible to have a
model
in which the time required for the occurrence of a singularity is infinite,
i.e., the singularity effectively does not happen.
However, if the energy density grows, the disintegration of bound structures
necessarily
has to take place, in a way similar to the case of a Big Rip.
Such Rip phenomena turns out to be common for Big Rip,  Little Rip or
Pseudo-Rip cosmologies, destroying all bound structures in a finite time. It is
remarkable that mild phantom scenarios like Little Rip or Pseudo-Rip may easily
mimic our current $\Lambda$CDM era and indicate to quite long existence
(billions of years!) of our universe before the Rip occurs.

The pioneering works on the future singularity \cite{caldwell02} were
considering the cosmic fluid to be non-viscous.
This is an idealized model, of course; it is useful in practice in many cases
but generally not able to
cope with intricate micro-scale phenomena that occur especially near solid
boundaries.
A next step in complexity is to allow for deviations from thermal equilibrium
to the first order.
That means, one  has to introduce two viscosity coefficients, namely the shear
viscosity $\eta$ and
the bulk viscosity $\zeta$.

In the present paper we consider dark energy era with  future Big Rip or Little
Rip
singularity under
a variety of conditions, assuming a more realistic form for the cosmic fluid
than what has
commonly been the case. We start in the next section by reviewing essentials of
the theory of a dark fluid satisfying the condition $p<-\rho$, i.e. a phantom
fluid. While in the general case one has to account for for both coefficients
$\eta$ and $\zeta$, as mentioned, we shall assume in conformity with usual
practice that spatial anisotropies are smoothed out. Thus only $\zeta$ has to
be included. Viscous Little Rip cosmology in an isotropic fluid has recently
been worked out; cf. Eq.~(\ref{7}) below. In Sec. III we focus attention on
the description of a {\it turbulent} state of the fluid
in the late universe. Such a possibility would seem physically most natural,
in view of the violent motions expected in the vicinity of the singularity.
We propose a two-component model in which, for physical reasons, the turbulent
energy component is put proportional to the scalar expansion. We consider in
this context two different proposals for the thermodynamic parameter
$w_{\rm turb}$ occurring in the equation of state. First, we put $w_{\rm turb}$
equal
to the usual parameter $w$ for non-turbulent matter, meaning that the turbulent
matter component behaves as a passive ingredient  as far as the equation of
state is concerned. Second, we allow for $w_{\rm turb}>-1$, meaning that we
cover the  region $-1<w_{\rm turb}<0$ also.
Both Big Rip and Little Rip evolutions for dark energy are found. In Sec. VI we
consider an approach
which is quite different, namely, we  model the universe as a one-component
dark fluid that becomes suddenly transformed into a turbulent state. A
noteworthy property of this model is that the turbulence in effect protects the
universe from encountering the future singularity at all. In Sec. VII we
present an equivalent description in terms of an inhomogeneous imperfect fluid,
including the effect of viscosity but not that of turbulence.
Some summary and outlook is given in Discussion section.

\section{Dark fluid with a bulk viscosity }

Consider the following model for the cosmic fluid in the later stages of the
development of the universe,
assuming that the fluid is dark and satisfies the inequality
\begin{equation}
p<-\rho\, ,
\label{1}
\end{equation}
so that the equation
\begin{equation}
\dot{H}=-\frac{1}{2}\kappa^2(\rho+p)
\label{2}
\end{equation}
with $H=\dot{a}/a$ and $\kappa^2=8\pi G$ implies the property
\begin{equation}
\dot{H}>0\, .
\label{3}
\end{equation}
This is the phantom region, corresponding to the thermodynamic parameter
$w=p/\rho$ being less than $-1$.
In the usual case of this type universe the future singularity is a true
mathematical singularity,
reached in a finite time, and is called the Big Rip. It is notable, however,
that in the limiting case
when $w \rightarrow -1$ from below, the future singularity is only
asymptotically reached.
This scenario is called the Little Rip.

It is noteworthy that in a cosmological context the value of $\eta$ appears to
be very much larger than
that of $\zeta$.
At least that is so in the earlier stages of the development of the universe
where the physical conditions
are better known than in the later stages  and conventional kinetic theory can
be used to calculate the viscosities.
A calculation of the viscosity coefficients was  made by Caderni and Fabbri
\cite{caderni77}.
For instance, considering the instant $t=1000$ s after the Big Bang, it turns
out that
\begin{equation}
\eta=2.8\times 10^{14} \mathrm{ g~ cm^{-1} s^{-1}}\, , \quad
\zeta=7.0 \times 10^{-3} \mathrm{ g ~cm^{-1} s^{-1}}\, ,
\label{4}
\end{equation}
(cf. also Ref.~\cite{brevik94}),
showing the large difference in magnitude between $\eta$ and $\zeta$.
Yet, it appears that spatial anisotropies in the universe are effectively
smoothed out, at least on a large scale,
so that in most current models the universe is assumed to be spatially
isotropic.
It means that  $\zeta$ is retained, but $\eta$ omitted in the Friedmann
equations.

A theory of viscous Little Rip cosmology was recently given in
Ref.~\cite{brevik11}.
Let us recapitulate  one of the characteristic results from that investigation:
If the effective pressure $p_\mathrm{eff}$ is  assumed  to have the explicit
form
\begin{equation}
p_\mathrm{eff}=-\rho-A\sqrt{\rho}-3\zeta H
\label{5}
\end{equation}
with $A$ a positive constant, and if moreover the bulk viscosity $\zeta$ is
assumed to satisfy the condition
\begin{equation}
3\zeta H \equiv \xi_0=\mathrm{constant}\, ,
\label{6}
\end{equation}
then the following expression is found for the time dependent energy density
\begin{equation}
\rho(t)=\left[ \left( \frac{\xi_0}{A}+\sqrt{\rho_0}\right)
\exp (\sqrt{6\pi G}\,At)-\frac{\xi_0}{A}\right]^2\, .
\label{7}
\end{equation}
Here subscript zero refers to the present time.
It is thus an infinite time needed to reach the infinite energy density case.
This is precisely the characteristic property of the Little Rip phenomenon.

\section{The turbulent approach}

Let us now apply a physical point of view on the dark energy universe, in its
later stages when it approaches the future singularity.
The simple description above, in terms of macroscopic bulk viscosity in the
fluid,
cannot be considered to be satisfactory, due to the following reason:
In the assumed  states of violent local fluid element motion a transition into
{\it turbulent} motion seems to be inevitable.
The local Reynolds number must be expected to be very high.
That brings, in fact, the {\it shear} back into the analysis, not in a
macroscopic sense as before,
but in a local sense causing  the distribution of local eddies over the wave
number spectrum.
What kind of turbulence should we expect? The natural choice is that of
isotropic turbulence,
which is a topic reasonably well understood. Thus, we should expect a
Loitziankii region for
low wave numbers where the energy density varies proportionally to $k^4$;
for higher $k$ we should expect an inertial subrange characterized by the
formula
\begin{equation}
E(k)=\alpha \epsilon^{2/3}k^{-5/3}
\label{8}
\end{equation}
with $\alpha$ the Kolmogorov constant and $\epsilon$ the mean energy
dissipation per unit time
and unit mass; and  finally when the values of $k$ become as high as the
inverse Kolmogorov length $\eta_L$,
\begin{equation}
k \rightarrow  k_L=\frac{1}{\eta_L}=\left(\frac{\epsilon}{\nu^3}\right)^{1/4}
\label{9}
\end{equation}
with $\nu$ the kinematic viscosity, we enter the dissipative region where the
local Reynolds number
is of order unity and heat dissipation occurs.
In accordance with common usage we shall consider the fluid system as
quasi-stationary,
and omit the production of heat energy. In practical cases it may be useful to
combine
these elements into the useful von K\'{a}rm\'{a}n interpolation formula which
covers the whole wave number spectrum (cf., for instance,
Refs.~\cite{panchev71,brevik92,carhart88}).

However, the full spectral theory of isotropic turbulence will no be needed in
our first approach
to the problem. Rather, we shall in the following focus attention on how the
turbulent part
of the energy density, called $\rho_\mathrm{turb}$, can be estimated to vary
from present time $t_0$ onwards.
First, we write the effective energy density  as a sum of two terms,
\begin{equation}
\rho_\mathrm{eff} = \rho + \rho_\mathrm{turb}\, ,
\label{10}
\end{equation}
where $\rho$ denotes the conventional macroscopic energy density in the local
rest inertial system of the fluid.
It is natural to assume that $\rho_\mathrm{turb}$ is proportional to $\rho$
itself.
Further, we shall assume that $\rho_\mathrm{turb}$ is proportional to the
scalar expansion
$\theta={U^{\mu}}_{;\mu}=3 H$.
This because physically speaking the transition to turbulence is expected to be
more pronounced
in the violent later stages, and a proportionality to the scalar expansion is
mathematically
the most simple way in which to represent the effect. Calling the
proportionality factor $\tau$,
we can thus  write the effective energy density as
\begin{equation}
\rho_\mathrm{eff}=\rho(1+3\tau H)\, . \label{11}
\end{equation}
Consider next the effective pressure $p_\mathrm{eff}$. We split it into two
terms,
\begin{equation}
p_\mathrm{eff}=p+p_\mathrm{turb}\, , \label{12}
\end{equation}
analogously as above.
For the conventional non-turbulent quantities $p$ and $\rho$ we assume the
standard relationship
\begin{equation}
p=w\rho\, ,
\label{13}
\end{equation}
where $-1<w<-1/3$ in the quintessence region and $w<-1$ in the phantom region.
The question now is: How does $p_\mathrm{turb}$ depend on $\rho_\mathrm{turb}$?
There seems to be no definite physical guidance to that problem, so we shall
make
the simplest possible choice in the following, namely write
\begin{equation}
p_\mathrm{turb}=w_\mathrm{turb}\,\rho_\mathrm{turb}\, , \label{14}
\end{equation}
with $w_\mathrm{turb}$ a constant.

We shall consider two different possibilities for the value of
$w_\mathrm{turb}$.
The first is to put $w_\mathrm{turb}$ equal to $w$ in Eq.~(\ref{13}), meaning
that the
turbulent matter behaves in the same way as the non-turbulent matter as far as
the equation
of state is concerned. This option is straightforward and natural,
and is not quite trivial since $\rho_\mathrm{turb}$ and $\rho$ behave
differently,
in view of Eq.~(\ref{11}).
Our second option will be to assume that $w_\mathrm{turb}$ takes another,
prescribed value.
In view of the expected violent conditions near the future singularity,
it might even be natural here to chose the value $w_\mathrm{turb}=+1$, i.e.,
the Zel'dovich fluid option.

The first and the second Friedmann equations  can now be written
\begin{equation}
H^2=\frac{1}{3}\kappa^2 \rho(1+3\tau H)\, ,
\label{15}
\end{equation}
\begin{equation}
\frac{2\ddot{a}}{a}+H^2=-\kappa^2  \rho(w+3\tau H w_\mathrm{turb}) \, .
\label{16}
\end{equation}
(recall that $\kappa^2=8\pi G$).
This may be compared with earlier attempt to introduce the turbulence in dark
energy \cite{brevik011}.

Equations (\ref{15}) and (\ref{16}) determine our physical model.
Recall that its input parameters are $\{w, w_\mathrm{turb}, \tau \}$, all
assumed constant.
   From these equations we can now describe the development of the Hubble
parameter.
For convenience we introduce the quantities $\gamma$ and
$\gamma_\mathrm{turb}$, defined as
\begin{equation}
\gamma=1+w, \quad \gamma_\mathrm{turb}=1+w_\mathrm{turb}\, .
\label{17}
\end{equation}
We can then write the governing equation for $H$ as
\begin{equation}
(1+3\tau H)\dot{H}+\frac{3}{2}\gamma H^2+\frac{9}{2}\tau
\gamma_\mathrm{turb}H^3=0\, .
\label{18}
\end{equation}
This equation is in principle to be integrated from present time $t=t_0=0$
onwards, with initial value $H=H_0=\dot{a}_0/a_0$.

\subsection{On the energy balance equation}

If $T_\mathrm{tot}^{\mu\nu}$ denotes the total energy-momentum tensor for the
cosmic fluid, we must have
\begin{equation}
{T_\mathrm{tot}^{\mu\nu}}_{;\nu}=0\, , \label{19}
\end{equation}
as a consequence of Einstein's equation.

In most cases studied, the expression for $T_\mathrm{tot}^{\mu\nu}$ can be
written down explicitly;
this is so for non-viscous fluids as well as with macroscopic viscous fluids.
In the present case this no longer true, however, since the turbulent energy is
produced by shear
stresses on a {\it small scale}, much less than the scale of the macroscopic
fluid equations.
That is, we are dealing with a non-closed physical system, of essentially the
same kind as encountered
in phenomenological electrodynamics in a continuous medium in special
relativity.
It implies that the source term in the energy balance equation has to be put in
by hand.

Let henceforth $T^{\mu\nu}$ refer to the non-viscous part of the fluid. We may
express the energy balance as
\begin{equation}
\dot{\rho}+3H(\rho+p)=-Q\, ,
\label{20}
\end{equation}
where the source term $Q$ is positive, corresponding to an energy sink for the
non-viscous fluid.
We shall put $Q$ equal to $\epsilon \rho$, where the specific energy
dissipation $\epsilon$ however
shall be taken to involve the large Hubble parameter $H$ in the later stage of
the development.
Let us assume the form
\begin{equation}
\epsilon=\epsilon_0(1+3\tau H)\, ,
\label{21}
\end{equation}
$\epsilon_0$ being the specific energy dissipation at present time.
This equation is seen to contain the same kind of development as assumed
before; cf. the analogous
Eq.~(\ref{11}). Thus, our ansatz for the energy balance reads
\begin{equation}
\dot{\rho}+3H(\rho+p) = -\rho \epsilon_0(1+3\tau H)\, .
\label{22}
\end{equation}
We shall in the following consider some examples.
First, however, it is of interest to compare Eq.~(\ref{22}) with the generic
equation commonly accepted
in order to deal with the extra {\it pressure} in a fluid,
\begin{equation}
T^{\mu \nu}=\rho U^\mu U^\nu +(p+\Pi)h^{\mu\nu}\, ,
\label{23}
\end{equation}
with $h^{\mu\nu}=g^{\mu\nu}+U^\mu U^\nu$ the projection tensor.
Here $\Pi$ is the extra pressure brought about by a variety of effects, among
them typically viscosity,
matter creation, or eventually a combination of these.
(For a more detailed discussion along the lines cf., for instance,
Refs.~\cite{calvao92,lima92,brevik96}.)
In the case of bulk viscosity, it is known that $\Pi=-3\zeta H$.
Thus, equation (\ref{22}) may be regarded as an energy equation analogous to
the pressure equation (\ref{23}).

\section{The case $w_\mathrm{turb}=w <-1$}

This case means that the turbulent component of the fluid is regarded as a
passive ingredient
as far as the equation-of-state (EoS) parameter is concerned.
The time development of $\rho$ and $\rho_\mathrm{turb}$ will however be
different.
Equation~(\ref{18}) reduces to
\begin{equation}
\dot{H}+\frac{3}{2}\gamma H^2=0\, ,
\label{24}
\end{equation}
leading to
\begin{equation}
H=\frac{H_0}{Z}\, ,
\label{25}
\end{equation}
where we have defined
\begin{equation}
Z=1+\frac{3}{2}\gamma H_0 t
\label{26}
\end{equation}
(note that $w<-1$ implies $\gamma <0$).
Thus we have a Big Rip cosmology, where the future singularity time $t_s$ is
given by
\begin{equation}
t_s=\frac{2}{3|\gamma|H_0}\, .
\label{27}
\end{equation}
The scale factor becomes correspondingly
\begin{equation}
a=a_0Z^{2/3\gamma}\, ,
\label{28}
\end{equation}
and from the first Friedmann equation (\ref{15}) we get the non-turbulent
energy density as
\begin{equation}
\rho=\frac{3H_0^2}{\kappa^2}\,\frac{1}{Z}\,\frac{1}{Z+3\tau H_0}\, .
\label{29}
\end{equation}
The ratio between turbulent and non-turbulent energy becomes
\begin{equation}
\frac{\rho_\mathrm{turb}}{\rho}=3\tau H=\frac{3\tau H_0}{Z}\, .
\label{30}
\end{equation}
It is of main interest to consider the behavior near $t_s$. As $Z=1-t/t_s$ we
see that
\begin{equation}
H \sim \frac{1}{t_s-t}\, , \quad a \sim \frac{1}{(t_s-t)^{2/3|\gamma|}}\, ,
\label{31}
\end{equation}
\begin{equation}
\rho \sim \frac{1}{t_s-t}, \quad \frac{\rho_\mathrm{turb}}{\rho} \sim
\frac{1}{t_s-t}\, .
\label{32}
\end{equation}
Notice the difference from conventional cosmology: the behavior of $H$ and $a$
near the singularity
is as usual, while the singularity of $\rho$ has become {\it weakened}.
The reason for this is, of course, the non-vanishing value of the parameter
$\tau$.
Moreover, as $t \rightarrow t_s$ all the non-turbulent energy has been
converted into turbulent energy.
   From a physical point of view, this is just as we would expect.

Let us examine how this formalism compares with our ansatz (\ref{22}) for the
energy balance
of the non-turbulent fluid.
The left hand side of that equation is explicitly calculable by means of the
formulas just derived.
It is convenient here to make use of the mathematical relationship
\begin{equation}
\dot{\rho}+3H(\rho+p)=a^{-3\gamma}\frac{d}{dt}(\rho a^{3\gamma})\, , \label{33}
\end{equation}
by means of which obtain
\begin{equation}
\dot{\rho}+3H(\rho+p)=
-\frac{27}{2}\frac{H_0^4|\gamma|\tau}{\kappa^2}\frac{1}{Z^2(Z+3\tau H_0)^2}\, .
\label{34}
\end{equation}
In the limit $t\rightarrow t_s$, this reduces to
\begin{equation}
\dot{\rho}+3H(\rho+p) \rightarrow -\frac{3}{2}\frac{H_0^2|\gamma|}{\kappa^2
\tau}\frac{1}{Z^2}\, .
\label{35}
\end{equation}
In Eq.~(\ref{22}) we calculate the right hand side
\begin{equation}
   - \rho \epsilon_0(1+3\tau
H)=-\frac{3H_0^2}{\kappa^2}\frac{\epsilon_0}{Z(Z+3\tau H_0)}
\left( 1+\frac{3\tau H_0}{Z}\right)\, ,
\label{36}
\end{equation}
which for $t\rightarrow t_s$ reduces to
\begin{equation}
   - \rho \epsilon_0(1+3\tau H) \rightarrow
   -\frac{3H_0^2}{\kappa^2}\frac{\epsilon_0}{Z^2}\, .
\label{37}
\end{equation}
This has actually the same form as Eq.~(\ref{35}), thus supporting the physical
consistency of the ansatz.
We obtain the relationship
\begin{equation}
\epsilon_0=\frac{1}{2}\frac{|\gamma|}{\tau}\, .
\label{38}
\end{equation}
This could hardly have been seen in advance.
The specific energy dissipation $\epsilon_0$ at the present time is related to
the EoS parameter $\gamma$
and the  parameter $\tau$ introduced in Eqs.~(\ref{11}) and (\ref{12}), in a
simple way.
In geometric units, the dimension of $\epsilon_0$ is cm$^{-1}$.

\subsection{A comment on Little Rip cosmology}

The theory given above concerns Big Rip, a singularity obtained in the future
at a finite time.
A milder variant of this is the Little Rip scenario, corresponding to an
infinite span of time
needed to reach the singularity (cf., for instance, Refs.~\cite{frampton11} and
\cite{brevik11}).
To achieve a Little Rip, we have to modify to some extent the above basic
assumptions.
One such modification is the following:
\begin{enumerate}
\item Take the equation of state to be
\begin{equation}
p=-\rho -A\sqrt{\rho}\, ,
\label{39}
\end{equation}
with $A$ a positive constant.
\item Put $\tau=0$ in the first Friedmann equation (\ref{15}).
\item Assume a milder form for the sink term in the energy balance equation
than the form given in (\ref{22}).
A simple option is to take $Q$ to be  proportional to $H$ itself.
Calling the proportionality constant $\xi_0~(>0)$, we  get the following
modified ansatz
\begin{equation}
\dot{\rho}+3H(\rho+p)=-\xi_0 H\, .
\label{40}
\end{equation}
   From Eqs.~(\ref{39}) and (\ref{40}), we now have
\begin{equation}
\dot{\rho}=(3A\sqrt{\rho}-\xi_0)H\, .
\label{41}
\end{equation}
Comparing this with the modified first Friedmann equation we get
\begin{equation}
t=\frac{\sqrt 3}{\kappa}\int_{\rho_0}^\rho
\frac{d\rho}{\sqrt{\rho}\,(3A\sqrt{\rho}-\xi_0)}
=\frac{2}{\sqrt 3}\frac{1}{\kappa A}\ln
\frac{3A\sqrt{\rho}-\xi_0}{3A\sqrt{\rho_0}-\xi_0} \, .
\label{42}
\end{equation}
Inverting this equation we have
\begin{equation}
\rho=\frac{\xi_0^2}{9A^2}\left[ 1+\left(\frac{3A\sqrt{\rho_0}}{\xi_0}-1\right)
\exp \left( \frac{1}{2}\sqrt{3}\,\kappa A t\right) \right]^2\, . \label{43}
\end{equation}
This expression shows just the characteristic property of the Little Rip
phenomenon:
the universe reaches the state $\rho \rightarrow \infty$, but needs an infinite
time to do so.
\end{enumerate}
One may ask: Are the physical assumptions underlying the Big Rip scenario
stronger than those underlying the Little Rip one?
In our opinion this is most likely so, although there are of course
considerable uncertainties
connected with the future time development of the universe.

\section{The  case $w<-1$, $w_\mathrm{turb}>-1$}

This case is thermodynamically quite different from the preceding one as the
turbulent component of the fluid is no longer a passive ingredient.
We have now $\gamma_\mathrm{turb}=1+w_\mathrm{turb} >0$,
which means that we  cover  the  region $-1<w_\mathrm{turb}<0$
also.
In the latter region, the turbulent
contribution to the pressure is still negative as above, while if
$w_\mathrm{turb}>0$
the turbulent pressure becomes {\it positive}, just as in ordinary
hydrodynamical turbulence.

The governing equation (\ref{18}), written as
\begin{equation}
(1+3\tau H)\dot{H}=\frac{3}{2}H^2(|\gamma|-3\tau \gamma_\mathrm{turb}H)\, ,
\label{44}
\end{equation}
tells us that at the present time $t=0$ the condition
\begin{equation}
|\gamma| > 3\tau \gamma_\mathrm{turb}H_0
\label{45}
\end{equation}
must hold.
This is so because at the present time, the turbulent part is regarded as
unimportant.
This corresponds to the inequality $\dot{H}>0$  (cf., Eq.~(\ref{3})).

Equation~(\ref{44}) can be integrated to give $t$ as a function of $H$,
\begin{equation}
t=\frac{2}{3|\gamma|}\left(\frac{1}{H_0}-\frac{1}{H}\right)-\frac{2\tau}{|\gamma|}
\left( 1+\frac{\gamma_\mathrm{turb}}{|\gamma|}\right)
\ln \left[ \frac{|\gamma|-3\tau \gamma_\mathrm{turb}H}{|\gamma|-3\tau
\gamma_\mathrm{turb}H_0}\frac{H_0}{H}\right] \, .
\label{46}
\end{equation}
A striking property of this expression is that it describes a {\it Little Rip}
scenario.
As $t\rightarrow \infty$, the Hubble parameter reaches a finite critical value
\begin{equation}
H_\mathrm{crit}=\frac{1}{3\tau}\frac{|\gamma|}{\gamma_\mathrm{turb}}\, .
\label{47}
\end{equation}
The physical role of $\gamma_\mathrm{turb}$ is thus to postpone and weaken
the development towards the future singularity.

A natural choice for the EoS parameter $w_\mathrm{turb}$ in the vicinity of the
singularity,
in view of the violent motions expected, would be
\begin{equation}
w_\mathrm{turb}=+1\, ,
\label{47A}
\end{equation}
that means, a Zel'dovich fluid. This is an extreme case,
where the velocity of sound equals the velocity of light.

\subsection{Vacuum non-turbulent fluid component ($w=-1$), $w_\mathrm{turb}>-1$
}

This case, corresponding to a vacuum non-turbulent fluid $(p=-\rho)$, has to be
considered
separately. From Eq.~(\ref{44}) we can solve directly for $H$ as a function of
$t$,
\begin{equation}
\frac{H}{H_0}=\frac{3\tau H_0+\sqrt{1+6\tau H_0+9\tau H_0^2(\tau
+\gamma_\mathrm{turb}t)}}{1+6\tau H_0+9\tau H_0^2\gamma_\mathrm{turb}t}\, .
\label{48}
\end{equation}
Thus as $t\rightarrow \infty$, $H$ decreases smoothly to zero as $t^{-1/2}$.
We conclude that at least quintessence conditions are necessary to produce a
future singularity.

\section{A one-component dark fluid}

We now turn to an approach that is quite different from
the one above,
namely to consider the cosmic fluid as a {\it one-component} fluid.
Thus the distinction between a non-turbulent and a turbulent fluid component is
avoided altogether.
This new approach is actually more close to the usual picture in hydrodynamics,
where
a fluid is known to shift suddenly from a laminar to a turbulent state.

Consider the following picture: the universe starts from present time $t=0$ as
an ordinary
viscous fluid with a bulk viscosity called $\zeta$, and develops
according to the Friedmann equations. We assume as before that the EoS
parameter $w<-1$,
meaning that the universe develops in the viscous era towards a future
singularity.
Before this happens, however, at some instant $t=t_*$, we assume that there is
a sudden transition of the whole fluid into a turbulent state, after which the
EoS parameter is $w_\mathrm{turb}$
and the pressure accordingly $p_\mathrm{turb}=w_\mathrm{turb}
\,\rho_\mathrm{turb}$.
As before we assume that $w_\mathrm{turb} >-1$, and for simplicity we take
$\zeta$, as well as $w$ and $w_\mathrm{turb}$, to be constants.
One may ask: What is the resulting behavior of the fluid, especially at later
stages?

The problem can easily be solved, making use of the condition that the density
of the fluid has to be continuous at $t=t_*$.
In the viscous era  $0<t<t_*$ the energy-momentum tensor of the fluid is
\begin{equation}
T_{\mu\nu}=\rho U_\mu U_\nu+(p-3H\zeta )h_{\mu\nu}\, ,
\label{49}
\end{equation}
where
\begin{equation}
h_{\mu\nu}=g_{\mu\nu}+U_\mu U_\nu \label{50}
\end{equation}
is the projection tensor (the shear viscosity is omitted because of spatial
isotropy).
Solving the Friedmann equations one gets \cite{brevik05,brevik10}
\begin{equation}
H=\frac{H_0\,e^{t/t_c}}{1-\frac{3}{2}|\gamma|H_0t_c(e^{t/t_c}-1)}\, ,
\label{51}
\end{equation}
\begin{equation}
a=\frac{a_0}{\left[1-\frac{3}{2}|\gamma|H_0t_c(e^{t/t_c}-1)
\right]^{2/3|\gamma|}}\, ,
\label{52}
\end{equation}
\begin{equation}
\rho=\frac{\rho_0\, e^{2t/t_c}}{\left[1-\frac{3}{2}|\gamma|H_0t_c(e^{t/t_c}-1)
\right]^2}\, ,
\label{53}
\end{equation}
where $t_c$ is the `viscosity time'
\begin{equation}
t_c=\left(\frac{3}{2}\kappa^2\zeta \right)^{-1}\, .
\label{54}
\end{equation}
The values $H_*, a_*, \rho_*$ at $t=t_*$ are thereby known.

In the turbulent era $t>t_*$ we can make use of the same expressions
(\ref{51}) - (\ref{53}) as above, only with substitutions
$t_c \rightarrow \infty~(\zeta \rightarrow 0)$, $t \rightarrow t-t_*,~ w
\rightarrow w_\mathrm{turb}$,
$H_0 \rightarrow H_*$, $a_0 \rightarrow a_*$, $\rho_0 \rightarrow \rho_*$. Thus
\begin{equation}
H=\frac{H_*}{1+\frac{3}{2}\gamma_\mathrm{turb}H_*(t-t_*)}\, ,
\label{55}
\end{equation}
\begin{equation}
a=\frac{a_*}{\left[ 1+\frac{3}{2}\gamma_\mathrm{turb}H_*(t-t_*)
\right]^{2/3\gamma_\mathrm{turb}}}\, ,
\label{56}
\end{equation}
\begin{equation}
\rho=\frac{\rho_*}{\left[ 1+\frac{3}{2}\gamma_\mathrm{turb}H_*(t-t_*)
\right]^{2}}
\label{57}
\end{equation}
(recall that $\gamma_\mathrm{turb} >0$).
Thus the density $\rho$, at first increasing with increasing $t$ according to
Eq.~(\ref{53}),
decreases again once the turbulent era has been entered, and goes smoothly
to zero as $t^{-2}$ when $t\rightarrow \infty$.
In this way the transition to turbulence protects the universe from entering
the future singularity.

It should be noted that whereas the density is continuous at $t=t_*$ the
pressure is not:
In the laminar era $p_*= w \rho_* <0$, while in the turbulent era $p_*
=w_\mathrm{turb}\,\rho_* $
will even be positive, if $w_\mathrm{turb}>0$.
Thus, we demonstrated the possible role of turbulence to protect the universe
from the future singularity. In the same fashion, one can consider its role in
protecting the universe from Rip, i.e. disintegration of bound structures.

\section{Inhomogeneous (imperfect) dark fluid description}

Let us present the equivalent formulation in terms of inhomogeneous (imperfect)
fluid. By including the effect by the viscosity but not including that by the
turbulence, we may consider the following EoS
\begin{equation}
\label{58}
p = - \rho + f(\rho) - 3\zeta H\, .
\ee
Here $f(\rho)$ is an appropriate function of the energy-density $\rho$.
The bulk viscosity $\zeta$ can be a function of $\rho$ and $H$,
$\zeta = \zeta\left(\rho, H\right)$.
Let assume the Hubble rate $H$ is given by a function $H=h(t)$.
Then by using the first FRW equation, we find
\begin{equation}
\label{59}
t = h^{-1}(H) = h^{-1} \left( \kappa \sqrt{\frac{\rho}{3}} \right)\, .
\ee
Then by using the second FRW equation, we obtain
\be
\label{60}
f(\rho) = - \frac{2}{\kappa^2}\dot H + 3 \zeta (\rho,H) H
= - \frac{2}{\kappa^2} h' \left( h^{-1} \left( \kappa \sqrt{\frac{\rho}{3}}
\right) \right)
+ \kappa \sqrt{3\rho} \zeta \left( \rho, \kappa \sqrt{\frac{\rho}{3}} \right)
\, .
\ee
Conversely, if $f(\rho)$ is given by (\ref{60}), we obtain a solution of the
FRW equation as $H=h(t)$.

Just for the simplicity, we may assume $3\zeta H$ is a constant as in
(\ref{6}),
\be
\label{61}
3 \zeta (\rho,H) H = \kappa \sqrt{3\rho} \zeta \left( \rho, \kappa
\sqrt{\frac{\rho}{3}} \right)
= \xi_0\, .
\ee
As a simple example, we may consider the model (\ref{25}) with (\ref{26}),
which is realized
by including the turbulence. Then we obtain
\be
\label{62}
f(\rho) = \gamma \rho + \xi_0\, .
\ee
As another example, we consider the model (\ref{44}), which give (\ref{45}).
Since
\be
\label{63}
h'(t) = \dot H = \frac{3H^2(|\gamma|-3\tau \gamma_\mathrm{turb}H)}{2(1+3\tau
H)}
= \frac{\kappa^2 \rho \left(|\gamma|- \tau \gamma_\mathrm{turb} \kappa
\sqrt{3\rho} \right)}{2\left(1+\tau \kappa \sqrt{3\rho} \right)}\, ,
\ee
we find
\be
\label{64}
f(\rho) = - \frac{\rho \left(|\gamma|- \tau \gamma_\mathrm{turb} \kappa
\sqrt{3\rho} \right)}{\left(1+\tau \kappa \sqrt{3\rho} \right)} + \xi_0 \, .
\ee
Then the models with turbulence can be equivalent to the inhomogeneous
(imperfect) fluid models without turbulence
with respect to the expansion history of the universe.
In other words, the effect of turbulence and/or of viscosity may be always
traded via the change of the effective equation of state.

\section{Discussion}

In summary, we have studied a dark fluid universe where a finite-time future Big Rip or a
Little Rip cosmology occurs. Special attention is paid to the role of viscosity
via its elementary physical properties. The possibility of a viscous Little Rip
cosmology where singularity occurs in the infinite future is confirmed. We
propose to take into account a turbulent state for the dark fluid in the late universe.
An explicit two-component model with  turbulent component proportional to the
Hubble rate is developed.
A conventional as well as a quintessence value for the turbulent component equation of
state parameter is considered, and the occurrence of a Big Rip and a Little Rip
cosmology is demonstrated. However, for a one-component fluid which suddenly
transforms into a turbulent state, the late-time acceleration universe is found to be
qualitatively different: the turbulence may protect the universe from encountering
a future singularity at all. It is interesting that an  equivalent description in terms
of an inhomogeneous fluid may  also be developed.

Turbulence is a fundamental property of classical fluids. It is known to be
important not only theoretically but also for number of practical applications
in everyday life (the air plane flight turbulence is a well known example).
The mysterious dark energy is often considered as some kind of
classical fluid with unusual properties.
It is then natural to expect (if such a picture is correct) that turbulence
phenomena may be important also for dark energy especially in the  very late violent
universe. In this work we have made a first step towards  the
construction of a bridge between turbulence and dark energy.
It turns out that turbulent dark energy may have quite rich properties, in
particular, in predicting the possible absence of future singularities.
   However, only forthcoming precise observational data about dark energy may confirm
the role of turbulence in the dark energy paradigm.

\section*{Acknowledgments.}

  SDO
has been partly supported by MICINN (Spain),  projects
FIS2006-02842 and FIS2010-15640, by the CPAN
Consolider Ingenio Project, by AGAUR (Generalitat de Ca\-ta\-lu\-nya),
contract 2009SGR-994 and by Eurasian National University.
SN is supported in part by Global COE Program of Nagoya University (G07)
provided by the Ministry of Education, Culture, Sports, Science \&
Technology and by the JSPS Grant-in-Aid for Scientific Research (S) \# 22224003
and (C) \# 23540296.


\end{document}